\documentclass[twocolumn,showpacs,preprintnumbers,amsmath,amssymb,prb]{revtex4}

\usepackage{graphicx}
\usepackage{dcolumn}
\usepackage{bm}

\def\rnum#1{\expandafter{%
\romannumeral #1}}
\def\Rnum#1{\uppercase\expandafter{%
\romannumeral #1}}

\begin{document}

\title{Coexistence of vector chiral order and Tomonaga-Luttinger liquid 
in the frustrated three-leg spin tube in a magnetic field}

\author{Masahiro Sato}

\affiliation{Synchrotron Radiation Research Unit, Japan Atomic 
Energy Agency, Sayo, Hyogo 679-5148, Japan\\
CREST Japan Science and
Technology Agency, Kawaguchi, Saitama 332-0012, Japan}

\date{\today}

\begin{abstract}
The frustrated three-leg antiferromagnetic spin-$\frac{1}{2}$ 
tube with a weak interchain coupling in a magnetic field 
is investigated 
by means of Abelian bosonization techniques.
It is clearly shown that a vector chiral long-range order 
and a one-component Tomonaga-Luttinger liquid coexist in a wide 
magnetic-field region from a state with a small magnetization 
to a nearly saturated one. 
The chiral order is predicted to still survive 
in the intermediate plateau state. 
We further predict that (even) when 
the strength of one bond in the three rung couplings 
is decreased (increased), an Ising-type quantum phase transition 
takes place and the chirality vanishes 
(no singular phenomena occur and the chiral order is maintained). 
Even without magnetic fields, the chiral order would also be present 
if the spin tube possesses easy-plane anisotropy.  
\end{abstract}


\maketitle

\section{\label{Intro} Introduction}
Quantum spin tubes$-$namely, systems of coupled spin chains with a 
{\it periodic boundary condition} (PBC) for the interchain (rung)
direction$-$have been studied since late
1990s.~\cite{Ka-Ta,Schu,Cab,Cit,Mill,Nojiri,Mila2,Mila3,O-Y,MS05NLSM,M-O} 
In particular, odd-leg spin tubes have attracted much interest, because the
antiferromagnetic (AF) rung coupling yields frustration. 
It has been recognized now that coupled spin chains including spin tubes
offer fairly richer physics compared with single spin chains.~\cite{texts} 
Recently, a few new spin-tube compounds 
such as a three-leg tube~\cite{Nojiri}
$\rm [(CuCl_2tachH)_3Cl]Cl_2$ and 
a nine-leg one~\cite{Mill} $\rm Na_2V_3O_7$ have been fabricated. 
The presence of these materials 
has further promoted the study of spin tubes.

For {\it nonfrustrated} AF spin-$S$ ladders$-$namely, 
coupled AF chains with an {\it open boundary condition} for the 
rung$-$the following universal feature 
(generalization of Haldane's prediction) 
has been elucidated:~\cite{Sier,Rojo,Dell} 
if both $2S$ and the number of chains, $N$, are odd, 
the low-energy physics is described by a gapless, 
one-component Tomonaga-Luttinger liquid (TLL) and 
no symmetry breakings take place, whereas for all the other cases
($2S\times N=$ even), the system has a gapped spectrum 
with conserving all symmetries.   
This even-odd property, however, does not always hold 
in {\it frustrated} coupled spin chains. 
In the three-leg AF spin-$\frac{1}{2}$ tube where the rung triangle 
induces geometrical frustration, 
it is known~\cite{Ka-Ta} that the one-site translational symmetry along 
the chain direction is spontaneously broken 
and a finite gap between a ground
state and the lowest excitation exists.  



In this paper, we point out another striking difference 
between the three-leg nonfrustrated ladder and the frustrated tube: 
namely, utilizing bosonization techniques, 
we 
show in a clear way that a vector chiral long-range order resides in the 
magnetic-field-induced TLL phase of the three-leg AF 
spin-$\frac{1}{2}$ tube at least in the weak-rung-coupling regime. 
In the chiral phase, the parity symmetry for the rung direction is
spontaneously broken. On the contrary, the chiral order is absent in the
corresponding ladder. 
For one-dimensional gapped $U(1)$-symmetric coupled spin chains, 
(\rnum{1}) the field-induced magnon-condensed
gapless state is usually described by a standard TLL and 
conserves all symmetries and 
(\rnum{2}) many physicists have focused only on whether 
nontrivial magnetization plateau states exist or not. 
Therefore, the possibility of chirality in spin tubes 
has not been considered well so far.~\cite{note_chiral} 
We also show that the chiral order continues in the intermediate
plateau state of the spin tube. These predictions means that the 
three-leg spin tube possesses both gapless and gapped 
chiral-ordered states. 
We further discuss a rung-coupling modification in the three-leg 
spin tube: the strength of one bond of 
three rung couplings is changed. If the strength of the changed bond 
is extremely increased (decreased), 
the system approaches a two-leg ladder plus single chain 
(a three-leg ladder). 
Both limits possess no geometrical frustration. 
Interestingly, it is found that in the case of increasing
the bond strength, the chiral order always remains unbroken, 
but an Ising-type quantum phase 
transition takes place and the chirality
disappears at the Ising critical point in the other case.

The organization of the paper is as follows. In Sec.~\ref{model}, we
define a spin tube model and briefly explain the bosonization
method. The next two sections are the main part of the paper. 
We show the mechanism of the vector chiral order in Sec.~\ref{analysis}. 
Section~\ref{rung-deform} is devoted to investigating effects of 
the rung deformation. Finally, conclusions and brief discussions 
about our results are presented in Sec.~\ref{conclusions}.

\section{\label{model} Model and Method} 
In this paper, we mainly consider the three-leg
AF spin-$\frac{1}{2}$ tube with
a homogeneous rung coupling. The Hamiltonian is defined as 
\begin{eqnarray}
\label{tube}
{\cal H} = \sum_{n=1}^3\sum_{j}\left[J\vec S_{n,j}\cdot\vec S_{n,j+1}
+J_\perp\vec S_{n,j}\cdot\vec S_{n+1,j}
-HS_{n,j}^z\right],
\end{eqnarray}
where $\vec S_{n,j}$ is spin-$\frac{1}{2}$ operator 
on site $j$ of the $n$th chain ($n=1$-$3$), $J>0$ ($J_\perp>0$) 
is the AF intrachain (rung) coupling, and the PBC 
$\vec S_{4,j}=\vec S_{1,j}$ is imposed. 
We begin with the independent three chains in a field $H$. 
From Abelian bosonization,~\cite{texts} 
the $n$th chain in the low-energy limit 
is mapped to a Gaussian model ${\cal H}_n = \int dx \,\,
\frac{v}{2}\left[
K(\partial_x\theta_n)^2+
K^{-1}(\partial_x\phi_n)^2\right]$,
where 
$\phi_n$ is the scalar boson field 
and $\theta_n$ is the dual to $\phi_n$. 
The TLL parameter $K$ and the
spin-wave velocity $v$ depend on $J$ and $H$; when $H$ is changed 
from $0$ to the saturation field $2J$, 
$v$ ($K$) monotonically decreases (increases) 
from $\pi J a_0/2$ ($1/2$) to $0$ ($1$) [$a_0$: lattice constant].  
The spin operators are also bosonized as 
\begin{subequations}
\label{spin_xxz}
\begin{eqnarray}
S_{n,j}^z&\approx& M+a_0 \partial_x{\phi_n}/\sqrt{\pi}
\nonumber\\
&&+(-1)^j A_1 \sin
(\sqrt{4\pi}{\phi_n}
+2\pi Mj)
+\cdots,
\end{eqnarray}
\begin{eqnarray}
S_{n,j}^+  &\approx& e^{i\sqrt{\pi}\theta_n}\big[(-1)^j B_0
\nonumber\\
&&+B_1\sin
(\sqrt{4\pi}{\phi_n}+2\pi M j)
+\cdots\big], 
\end{eqnarray}
\end{subequations}
where $M(H)=\langle S_{n,j}^z\rangle$, and 
$A_l$ and $B_l$ are nonuniversal constants of $O(1)$.~\cite{Lu-Za,Hi-Fu} 
This formula indicates that the period of $\phi_n$ ($\theta_n$) 
is $\sqrt{\pi}$ ($\sqrt{4\pi}$).
Using ${\cal H}_n$ and formula~(\ref{spin_xxz}), we can
straightforwardly derive the following effective Hamiltonian 
of the model~(\ref{tube}) in the weak-rung-coupling regime: 
\begin{eqnarray}
\label{eff_1}
{\cal H}_{\rm eff} = \int dx \,\,
\sum_{n=1}^3 \frac{v}{2}\left[
K(\partial_x\theta_n)^2+K^{-1}
(\partial_x\phi_n)^2\right]
\nonumber\\
+ \frac{2MJ_\perp}{\sqrt{\pi}}\sum_n\partial_x\phi_n 
+\frac{J_\perp a_0}{\pi}\sum_n\partial_x\phi_n \partial_x\phi_{n+1}
\nonumber\\
+\sum_n J_\perp a_0^{-1}\Big[B_0^2\cos(\sqrt{\pi}(\theta_n-\theta_{n+1}))
\nonumber\\
+\frac{A_1^2}{2}\cos(\sqrt{4\pi}(\phi_n-\phi_{n+1}))
\Big]+\cdots,
\end{eqnarray}
where $\phi_4=\phi_1$, $\theta_4=\theta_1$, and we have written only
the important part among all the rung-coupling terms and neglected 
terms with oscillating factors $e^{li\pi Mj}$ or $(-1)^j$, 
which are irrelevant. 
In the bosonization picture, symmetries of the spin tube~(\ref{tube})
are represented as follows: a $U(1)$ rotation around the $S^z$ axis 
$S^+_{n,j}\to e^{i\gamma}S^+_{n,j}$, the one-site translation 
along the chain $S^\alpha_{n,j}\to S^\alpha_{n,j+1}$, 
that along the rung $S^\alpha_{n,j}\to S^\alpha_{n+1,j}$, 
the site-parity transformation for the chain 
$S^\alpha_{n,j}\to S^\alpha_{n,-j}$, and that for the rung 
$S^\alpha_{1,j}\leftrightarrow S^\alpha_{3,j}$, respectively, correspond to 
$\theta_n\to\theta_n+\gamma/\sqrt{\pi}$, 
$(\phi_n,\theta_n)\to(\phi_n+\sqrt{\pi}(M+1/2),\theta_n+\sqrt{\pi})$, 
$(\phi_n,\theta_n)\to(\phi_{n+1},\theta_{n+1})$,
$(\phi_n(x),\theta_n(x))\to(-\phi_n(-x)+\sqrt{\pi}/2,\theta_n(-x))$ 
and $(\phi_1,\theta_1)\leftrightarrow(\phi_3,\theta_3)$. 
These symmetries strongly restrict possible terms in ${\cal H}_{\rm eff}$. 
In other words, the symmetries reduce the number of coupling constants.  
As a result, for all vertex operators without oscillating factors and
conformal spin, 
only $\sum_n\cos[l\sqrt{4\pi}(\phi_n-\phi_{n+1})]$ and 
$\sum_n\cos[l\sqrt{4\pi}(\theta_n-\theta_{n+1})]$ are allowed to exist in
Eq.~(\ref{eff_1}). 
The most relevant terms with $l=1$ indeed appear in Eq.~(\ref{eff_1}). 
Utilizing the effective theory~(\ref{eff_1}), 
we study the low-energy properties of the spin tube below. 
We 
stress here that almost all the discussions below are applicable
to other three-leg spin tubes with 
symmetry-conserved perturbations 
($XXZ$ anisotropy, further-neighbor interactions, etc.). 

\section{\label{analysis} Vector Chiral Order}
\begin{figure}
\scalebox{0.5}{\includegraphics[width=\linewidth]{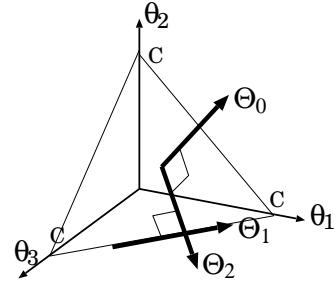}}
\caption{\label{new_boson} Relationship between $\{\theta_n\}$ 
and $\{\Theta_q\}$. $C$ is a constant.}
\end{figure}

Let us introduce new boson fields  
$\Phi_0=\sum_n\phi_n/\sqrt{3}$, $\Phi_1=(\phi_1-\phi_3)/\sqrt{2}$, 
$\Phi_2=(\phi_1+\phi_3-2\phi_2)/\sqrt{6}$, 
$\Theta_0=\sum_n\theta_n/\sqrt{3}$, 
$\Theta_1=(\theta_1-\theta_3)/\sqrt{2}$, and 
$\Theta_2=(\theta_1+\theta_3-2\theta_2)/\sqrt{6}$. The relationship
between old and new bosons is illustrated in Fig.~\ref{new_boson}. 
As one will see later, these fields help us to detect a 
vector chiral order. 
Using these new fields, we can diagonalize the boson 
bilinear part in Eq.~(\ref{eff_1}). Consequently, the effective 
Hamiltonian is rewritten as 
\begin{eqnarray}
\label{eff_2}
{\cal H}_{\rm eff} &=& \int dx \,\,
\sum_{q=0}^2 \frac{v_q}{2}\left[
K_q(\partial_x\Theta_q)^2+
K_q^{-1}(\partial_x\Phi_q)^2\right]
\nonumber\\
&&
+ 2\sqrt{\frac{3}{\pi}}MJ_\perp \partial_x\Phi_0 
+B_0^2\frac{J_\perp}{a_0} V[\Theta_1,\Theta_2]
\nonumber\\
&&+A_1^2 \frac{J_\perp}{2a_0} V[\Phi_1,\Phi_2]+\cdots,
\end{eqnarray}
where the potential $V[\alpha,\beta]$ is defined as
\begin{eqnarray}
\label{potential}
V = 2\cos
\Big(\sqrt{\frac{\pi}{2}}\alpha\Big)
\cos
\Big(\sqrt{\frac{3\pi}{2}}\beta\Big)
+\cos\left(\sqrt{2\pi}\alpha\right).
\end{eqnarray}
The new TLL parameters $K_q$ and velocities $v_q$ are evaluated as 
\begin{eqnarray}
\label{K_and_v}
K_0=K f_0^{-1},&& v_0=v f_0,
\nonumber\\
K_{1,2}=K f_g^{-1}=K_g, &&  v_{1,2}=v f_g=v_g,
\end{eqnarray}
where $f_0=(1+\frac{2K}{\pi}\frac{J_\perp a_0}{v})^{1/2}$ and 
$f_g=(1-\frac{K}{\pi}\frac{J_\perp a_0}{v})^{1/2}$. 
One finds that the Hamiltonian~(\ref{eff_2}) does not contain any
vertex operator with $\Phi_0$ or $\Theta_0$. This is strongly
supported by symmetries of the tube. The $\Phi_0$ sector hence provides
a one-component TLL. The derivative term $\partial_x\Phi_0$ can be
absorbed into the Gaussian part and yields only a small correction
of $M$. On the other hand, the remaining $\Phi_{1,2}$
sectors are subject to the effect of the potential $V$.  
Since the scaling dimensions of $V[\Phi_1,\Phi_2]$ and
$V[\Theta_1,\Theta_2]$ are, respectively, evaluated as $2K_g$
and $1/(2K_g)$, the latter $V[\Theta_1,\Theta_2]$ is always dominant and
relevant [i.e., $1/(2K_g)<2$] from a small $M$ to the saturation.  
The phase fields $\Theta_1$ and $\Theta_2$ thus are pinned at the
minimum of $V[\Theta_1,\Theta_2]$, and $\Phi_{1,2}$ sectors have
massive spectra. Consequently, the low-energy physics of the
tube~(\ref{tube}) is governed by a TLL in the $\Phi_0$ sector. 
The TLL just corresponds to the known field-induced
critical phase.~\cite{Cab} 
[For the case of $M=0$, $e^{li\pi Mj}=1$ and 
relevant vertex operators with $\Phi_0$ appear in Eq.~(\ref{eff_2}). 
They are expected to cause the $\Phi_0$ sector to obtain a gapped spectrum.]

\begin{figure}
\scalebox{0.95}{\includegraphics[width=\linewidth]{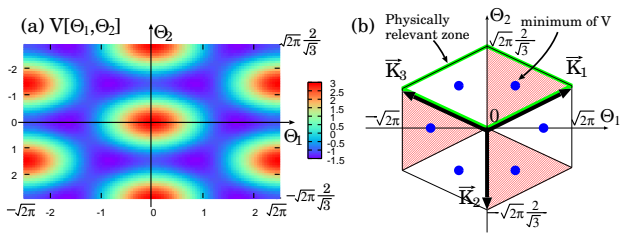}}
\caption{\label{potential_zone} (a) Potential $V[\Theta_1,\Theta_2]$. 
(b) Physically relevant zone (enclosed by the green line) 
in the projected $\Theta_1$-$\Theta_2$ plane. 
$\vec K_n$ play the role of the primitive translation vectors.  
The rung-parity transformation causes $\Theta_1\to-\Theta_1$, 
and the rung translation by one site does $\vec K_n\to\vec K_{n+1}$.}
\end{figure}

The scenario leading to the TLL here is already known well.~\cite{OYA,Cab} 
However, as one will see below, characteristics of the
frustrated spin tube~(\ref{tube}) are 
hidden in the form of the potential $V$.  
Figure~\ref{potential_zone}(a) and Eq.~(\ref{potential}) show that 
six minimum points 
$(\Theta_1,\Theta_2)=(\pm\sqrt{2\pi}/3,\pm\sqrt{2\pi/3})$ and
$(\pm2\sqrt{2\pi}/3,0)$ lie around the origin $(0,0)$. 
[Inversely, $(0,0)$ is the unique minimum of the potential 
$J_\perp V$ for $J_\perp<0$.]
Not all, however, are physically meaningful, because phase fields 
$\theta_n$ have a period. Projecting the physically relevant cubic space
$\theta_n\in[-\sqrt{\pi},\sqrt{\pi}]$ (mod $\sqrt{4\pi}$) 
onto the $\Theta_1$-$\Theta_2$ plane (see Fig.~\ref{new_boson}), 
we obtain a diamond 
zone as in Fig.~\ref{potential_zone}(b). 
Therefore, 
it is enough to consider only two minimum points 
$(\Theta_1^\pm,\Theta_2)=(\pm\sqrt{2\pi}/3,\sqrt{2\pi/3})$ that 
satisfy $\Theta_1^+=-\Theta_1^-\neq0$. 
Since the rung-parity transformation
$(\phi_1,\theta_1)\leftrightarrow(\phi_3,\theta_3)$ causes 
$\Theta_1\to-\Theta_1$, pinning $\Theta_1$ to these two minima
implies the spontaneous breakdown of the rung-parity symmetry. 
As a candidate of the rung-parity order parameter, we can propose a
vector chirality $\vec\kappa_{n,j}=\vec S_{n,j}\times\vec S_{n+1,j}$,
which changes sign via the rung-parity operation. 
From the formula~(\ref{spin_xxz}), the z component of a 
chirality is evaluated as 
\begin{eqnarray}
\label{chiral}
\langle \kappa_{3,j}^z \rangle 
\approx -B_0^2 
\left\langle\sin\Big(\sqrt{2\pi}\Theta_1\Big)\right\rangle
+\cdots= \mp \,{\rm finite} ,
\end{eqnarray}
for $\Theta_1^\pm$. Remarkably, the leading term of the chirality 
does not contain the massless fields $(\Phi_0,\Theta_0)$. 
Similarly, $\langle \kappa_{1(2),j}^z \rangle$ are shown to be 
equivalent to $\langle \kappa_{3,j}^z \rangle$. 
We thus conclude that the vector chiral long-range order exists in the
field-induced TLL phase. The chirality correlation function 
exhibits an exponential decay: 
$\langle \kappa_{3,j}^z\kappa_{3,0}^z \rangle\approx 
\langle\kappa_{3,j}^z\rangle^2+C_0 e^{-|j|/\xi}/\sqrt{|j|}+\cdots$.
Since any artificial approximation such as a mean-field decoupling 
is not used to obtain Eq.~(\ref{chiral}), 
the prediction of the chiral order is highly reliable. 
Note that in the chiral phase, the spontaneous breakdown of 
the $Z_2$ rung-parity symmetry occurs, but the $U(1)$ spin-rotational
symmetry is preserved.

For the spin-$\frac{1}{2}$ tube~(\ref{tube}), it is predicted that the 
intermediate plateau state appears at $M=1/6$ in the region 
$J_\perp>J_\perp^c(\sim 0.1J)$.~\cite{Cab} 
In the bosonization picture, the spin gap of the plateau state is
interpreted as that of the $\Phi_0$ sector, which is induced by a relevant
higher-order umklapp term $\cos(4\sqrt{3\pi}\Phi_0+12\pi Mj)
=\cos(4\sqrt{3\pi}\Phi_0)$. On the other hand, 
the $\Phi_{1,2}$ sectors are hardly influenced by whether the
$\Phi_0$ sector obtains a gapped spectrum or not. Therefore, we predict
that the chiral order still continues in the intermediate plateau phase
at least if $J_\perp(>J_\perp^c)$ is sufficiently small. 
This chiral plateau state is reminiscent of the narrow chiral phase in
the classical AF $XY$ model on a triangular lattice.~\cite{Mi-Sh} 
It is noteworthy that by controlling the strength of the magnetic
field, one can realize both {\it gapless and gapped} chiral states in
{\it single} spin tube.

\section{\label{rung-deform} Rung Deformation}
\begin{figure}
\scalebox{0.8}{\includegraphics[width=\linewidth]{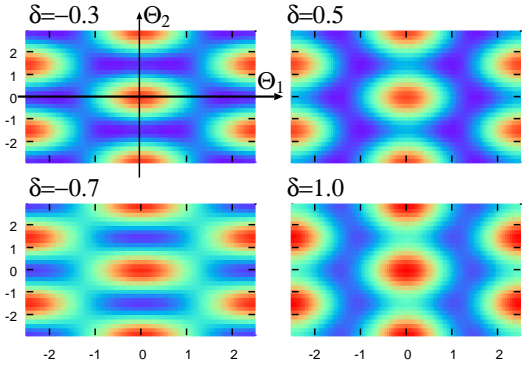}}
\caption{\label{deformation} Deformed potential 
$V_\delta[\Theta_1,\Theta_2]$.}
\end{figure}

In this section, we discuss the rung deformation. 
Let us modify the coupling between the first and third chains as 
\begin{eqnarray}
\label{rung_modify}
J_\perp\vec S_{1.j}\cdot\vec S_{3,j}&\to&
J_\perp(1+\delta)\vec S_{1.j}\cdot\vec S_{3,j}.
\end{eqnarray}
This explicitly violates the rung translational symmetry,
but conserves the parity symmetry between the first and third chains. 
Therefore, $\langle \kappa_{3,j}^z \rangle$ is still valid as a
rung-parity order parameter. The cases of $\delta=-1$ and 
$\delta\to+\infty$, respectively, correspond to a three-leg ladder and
a system of a two-leg ladder plus single chain. 
Note here that if we set $J_\perp\delta={\rm const}\ll J$, 
the present weak-rung-coupling approach is available even for 
the case of a large $|\delta|\gg 1$.  
A finite $\delta$ brings new bosonic terms 
$\delta J_\perp M\partial_x(\phi_1+\phi_3)/\sqrt{\pi}$, 
$\delta J_\perp a_0\partial_x\phi_1\partial_x\phi_3/\pi$, 
$\delta J_\perp B_0^2 a_0\cos(\sqrt{2\pi}\Theta_1)$, etc. 
The first term can be absorbed into the Gaussian part via the shift of
$\Phi_q$, which does not affect dual fields $\Theta_q$ and yields
a small deviation from the uniform magnetization, 
$\langle S_{2,j}^z\rangle\neq\langle S_{1(3),j}^z\rangle$. 
It is not surprising because (as mentioned above) 
$\delta$ breaks the rung translational symmetry. 
The second boson bilinear term is expected not to qualitatively 
influence the low-energy physics. 
Actually, introducing a new basis different from $(\Phi_q,\Theta_q)$, 
we can diagonalize all the bilinear terms. 
The main effect of the rung modification originates
from the third cosine term which varies the form of the potential
$V[\Theta_1,\Theta_2]$ as follows: 
\begin{eqnarray}
\label{modified_pot}
V[\Theta_1,\Theta_2]\to
V_\delta[\Theta_1,\Theta_2]=\hspace{3cm}
\\
2\cos
\Big(\sqrt{\frac{\pi}{2}}\Theta_1\Big)
\cos
\Big(\sqrt{\frac{3\pi}{2}}\Theta_2\Big)
+(1+\delta)\cos\left(\sqrt{2\pi}\Theta_1\right).\nonumber
\end{eqnarray}
The deformed potential $V_\delta$ is illustrated in
Fig.~\ref{deformation}.  
From $V_\delta$, one finds that when $\delta$ increases 
[decreases and reaches $\delta_{\rm cl}=-0.5$], 
the two minimum points still survive with keeping $\Theta_1\neq 0$ 
[approach with each other and meet at
$(\Theta_1,\Theta_2)=(0,\sqrt{2\pi/3})$]. The chirality 
$\langle \kappa_{3,j}^z \rangle$
is hence shown to be always finite when an arbitrary positive 
$\delta$ is applied. In other words, once an infinitesimal exchange is
added between a two-leg AF ladder and single chain, a vector chiral order
immediately emerges. On the other hand, 
for the case of $\delta<0$, the vector chiral order is
predicted to vanish at a certain point $\delta=\delta_c$ and the
low-energy physics is described by the TLL of the $\Phi_0$ sector for
$\delta<\delta_c$. 
The true value of $\delta_c$ must be renormalized from 
the classical value $\delta_{\rm cl}$ 
by the effect of the Gaussian part 
and must depend on $J_\perp/J$ and $H/J$.

In order to further understand the physics near the phase transition at
$\delta=\delta_c$, it is necessary to go beyond the semiclassical
analysis of $V_\delta$. 
If $\cos(\sqrt{3\pi/2}\Theta_2)$ in $V_\delta$ 
is allowed to be replaced with its expectation value, 
the effective Hamiltonian for the $\Phi_1$ sector is written as the following
double sine-Gordon ($d$SG) model~\cite{doubleSG1,doubleSG2}: 
\begin{eqnarray}
\label{doubleSGmodel}
{\cal H}_{[\Phi_1,\Theta_1]}\approx 
\int dx \, \left[K_g(\partial_x\Theta_1)^2
+K_g^{-1}(\partial_x\Phi_1)^2\right]\hspace{0.5cm}
\nonumber\\
-\frac{J_\perp}{a_0} B_0^2
\left[C_1 \cos\big(\sqrt{\frac{\pi}{2}}\Theta_1\big)
-(1+\delta)\cos\big(\sqrt{2\pi}\Theta_1\big)\right],
\end{eqnarray}
where $C_1=\langle \cos(\sqrt{3\pi/2}\Theta_2)\rangle$. 
It is believed that for $K_g>1/4$, 
the $d$SG model exhibits a second-order Ising transition by tuning
the coupling constants. 
Indeed, as $\delta$ passes through
a critical value $\delta_c$, the potential of the $d$SG model changes
from a double-well type to a single-bottom one. 
We thus conclude that the transition at $\delta=\delta_c$ 
is in the Ising universality class. The semiclassical analysis of $V_\delta$ 
becomes more reliable with $V_\delta$ being more relevant. 
It is hence expected that the value $\delta_c$ 
monotonically decreases and approaches $\delta_{\rm cl}$ 
when $K_g(\approx K)$ increases$-$i.e., 
when $H$ becomes close to the upper critical field.
Equation~(3.11) and Fig.~12 in Ref.~\onlinecite{doubleSG2} also
support this prediction.

Since $\langle \kappa_{3,j}^z \rangle\propto
\langle\sin(\sqrt{2\pi}\Theta_1)\rangle 
\approx \sqrt{2\pi} \langle \Theta_1\rangle$ 
holds near the transition, the chirality would play the role of 
the Ising order parameter. 
Comparing our $d$SG model~(\ref{doubleSGmodel}) and results [Eqs.~(29) and
(58)] in Ref.~\onlinecite{doubleSG1}, we can find
\begin{eqnarray}
\label{Ising_critical}
\langle \kappa_{3,j}^z \rangle \sim (\delta-\delta_c)^{1/8},
\end{eqnarray}
near $\delta\to \delta_c+0$. 
[The mean-field analysis of $V_\delta$ leads to 
$\langle \kappa_{3,j}^z \rangle \sim (\delta-\delta_c)^{1/2}$.] 
The chirality correlation function behaves as 
$\langle \kappa_{3,j}^z \kappa_{3,0}^z \rangle\sim C_2/|j|^{1/4}$ just at the
critical point. Moreover, near the point, it follows the results in
Ref.~\onlinecite{McWu}.       

\begin{figure}
\scalebox{0.7}{\includegraphics[width=\linewidth]{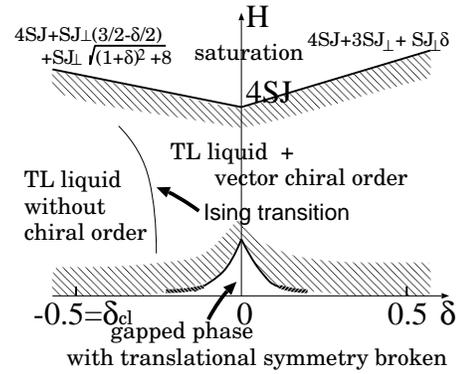}}
\caption{\label{phases} Phase diagram of the spin-$\frac{1}{2}$ tube. 
$J_\perp(<J_\perp^c)$ is fixed in the weak-rung-couping regime. 
The bosonization approach is less valid near both the lower and the 
upper critical fields. The upper critical field here is evaluated 
by the energy dispersion of one magnon.}
\end{figure}

Sakai {\it et al}. have recently predicted 
that when $\delta$ is added to the spin tube in
{\it zero field}, the spin gap rapidly disappears.~\cite{Sakai} 
Combining this prediction and ours, 
we construct the phase diagram on the
$\delta$-$H$ plane as in Fig.~\ref{phases}.

\section{\label{conclusions} Conclusions and Discussions} 
We have studied the three-leg AF spin-$\frac{1}{2}$ tube~(\ref{tube}) 
with a weak rung coupling under a magnetic field 
making use of Abelian bosonization techniques. 
In Sec.~\ref{analysis}, we have definitely elucidated without the help
of any artificial approximation 
that the vector chiral long-range order occurs in 
the magnetic-field-driven one-component TLL phase. 
Due to the chiral order, 
the discrete rung-parity symmetry is spontaneously broken, but the
$U(1)$ spin-rotational symmetry remains unbroken. 
We have also been predicted that the chiral order survives even in
the intermediate plateau state with $M=1/6$. 
It is remarkable that one can obtain both gapless and gapped 
chiral states by changing the strength of the magnetic field.  
In Sec.~\ref{rung-deform}, we have investigated effects of the rung
deformation in Eq.~(\ref{rung_modify}). It has been shown that 
when a positive $\delta$ is introduced with fixing $J_\perp\delta$ 
the chiral order and the TLL always continue, whereas when $\delta$ is 
decreased the chiral order vanishes at a certain Ising critical point
$\delta=\delta_c<0$ and then a standard TLL phase appears
(see Fig.~\ref{phases}). This implies that vector chirality 
correlations are favored in two-leg spin ladders~\cite{2leg_sato} 
as well as in three-leg tubes.

Quite recently, based on the spin-wave picture, 
we have predicted that the same chiral order still survives a
certain regime in the vicinity of saturation of the spin tube with 
$\delta=0$.~\cite{SW_sato} 
Therefore, it is inferred that the tube possesses 
the chiral order in a quite wide range of the magnetic field. 
One immediately finds that the classical three-leg spin tube always has
a vector chiral order from the zero field to the upper critical one.  
We thus may say that the classical nature is strong in the
weak-rung-coupling area of the quantum spin tube~(\ref{tube}). 
Actually, as well known, the AF spin-$\frac{1}{2}$ chain, 
which is the starting point of our analysis, 
has a large instability toward a classical N\'eel ordering.

The discussion in this paper tells us that 
even without magnetic fields, a vector chiral order could also occur 
if a spin tube consists of three spin chains with a large $K(>1/2)$:
for instance, the condition $K>1/2$ is satisfied in a chain with 
easy-plane $XXZ$ anisotropy. It is predicted that chiral orders 
exist in a certain parameter regime of 
easy-plane zigzag spin chains.~\cite{Hi-Ka} 
Our results also suggests that a field-induced 
vector chiral order can appear in other frustrated odd-leg spin 
tubes$-$i.e., five-, seven-, nine-leg tubes, etc., tubes.

Finally, we briefly mention the possibility of experimentally 
detecting the vector chirality. The chiral order predicted 
in this paper must be destroyed by an effect of thermal fluctuation. 
However, chirality correlations are expected to still 
be strong if the temperature is sufficiently low.   
Moreover, weak three-dimensional interactions among tubes can stabilize
the long-range chiral order: the gapless chiral order in the tube
is expected to change into a conventional umbrella spin structure 
due to the interactions. Four-point spin correlation functions include 
features of the above chiral order or strong chiral correlation. 
Such functions, in principle, are measured 
in polarized neutron scattering, electromagnetic-wave resonance, etc.

\begin{acknowledgments}
This work is supported by a Grant-in-Aid for Scientific 
Research (B) (No. 17340100) from the Ministry of Education, 
Culture, Sports, Science and Technology of Japan.
\end{acknowledgments}



\bibliography{apssamp}

\end{document}